\documentclass[conference]{IEEEtran}
\usepackage{graphicx}
\usepackage{cite}
\usepackage{tabularx}
\usepackage{makecell}
\usepackage{booktabs} 
\usepackage{tabularray}
\usepackage{orcidlink}
\usepackage{multirow} 
\usepackage{booktabs} 
\usepackage{caption}  
\usepackage{array}
\usepackage{hyperref}
\usepackage{float}
\usepackage{placeins}
\usepackage{xurl}

\title{Quantum Readiness in Latin American High Schools: Curriculum Compatibility and Enabling Conditions}

\author{
    \IEEEauthorblockN{Adriana Celeste Alvarado León}\orcidlink{0009-0002-3182-7708}
    \IEEEauthorblockA{Universidad de Ingeniería y Tecnología \\
    Lima, Perú \\
    Email: adriana.alvarado@utec.edu.pe}
    \and
    \IEEEauthorblockN{Osmar Denilson Herrera Cueva}\orcidlink{0009-0002-8115-0803}
    \IEEEauthorblockA{Colegio Trilce Los Olivos\\
    Lima, Perú \\ 
    Email: 73549107@trilce.edu.pe}
    \and
    \IEEEauthorblockN{Rosario Mercedes Morales Orvezo}\orcidlink{0009-0008-2580-1679}
    \IEEEauthorblockA{Colegio Mayor Secundario Presidente del Perú \\
    Lima, Perú \\
    Email: 72602393@lima.coar.edu.pe}
    \and
        \IEEEauthorblockN{Daniella Alexandra Crysti Vargas Saldaña}\orcidlink{0009-0001-6926-2469}
    \IEEEauthorblockA{Universidad Peruana de Ciencias Aplicadas \\
    Lima, Perú \\
    Email: daniellavargas@ieee.org}
    \and
            \IEEEauthorblockN{Freddy Herrera Cueva}
    \orcidlink{0009-0009-4793-7195}
    \IEEEauthorblockA{University of Maryland\\
    Maryland, USA \\
    Email: fredhc25@umd.edu}
}

\begin{document}

\maketitle

\begin{abstract}
The accelerating global development of quantum technologies strengthens the case for introducing quantum computing concepts before university. Yet in Latin America, there is no consolidated, region-wide integration of quantum computing into secondary education, and the feasibility conditions for doing so remain largely unexamined. This paper proposes a qualitative, comparative framework to assess academic readiness for quantum education across six countries—Peru, Bolivia, Chile, Argentina, Brazil, and Colombia—grounded in the relationship between curriculum compatibility and enabling conditions spanning institutional capacity, teacher preparation, infrastructure, and equity. Using official curricula, policy documents, national statistics, and educational reports, we apply structured qualitative coding and a 1–5 ordinal scoring system to generate a cross-country diagnosis. The findings reveal substantial regional asymmetries: among the six countries studied, Chile emerges as the most institutionally prepared for progressive quantum-education integration, while the remaining countries exhibit varying combinations of curricular gaps and fragmented but promising enabling conditions. Building on this diagnosis, we propose a country-sensitive, regionally coordinated roadmap for staged implementation—beginning with teacher development and pilot centers, leveraging open-source platforms and local-language resources, and scaling toward gradual curricular integration. This work establishes a baseline for future quantitative and mixed-method studies evaluating learning outcomes, motivation, and scalable models for quantum education in Latin America. 
\end{abstract}

\section{Introduction}
Quantum computing has emerged as a field of increasing global relevance, with the power to revolutionize how complex problems are tackled in science, industry, and everyday life. In recent years, this interest has begun to spread to the educational sphere, with initiatives aimed at introducing 
quantum concepts at the pre-university level.

However, existing literature on quantum education has largely focused on global frameworks, pilot initiatives, or high-income contexts, proposing generic curriculum models. While valuable, these approaches tend to overlook contextual constraints and enabling factors that condition feasibility in developing countries. To date, no study has offered a region-specific analysis of the educational, institutional, and socioeconomic conditions that determine the readiness of Latin American high school educational systems for quantum computing.

This article addresses this gap by introducing a qualitative and comparative framework for assessing academic readiness for quantum computing education in Latin America. Academic readiness is conceptualized as a multidimensional concept that encompasses curriculum coverage, alternative educational pathways, teacher training, technological infrastructure, institutional and political support, socioeconomic conditions, and student culture and participation. Rather than directly assessing students' performance, this study examines the structural and systemic conditions that enable or limit the introduction of quantum-related content at the high school level.

Using data from official curricula, national statistics, policy documents, and educational programs, we apply this framework to a qualitative analysis of selected Latin American countries. The results provide a detailed diagnosis of strengths, limitations, and asymmetries across the region, highlighting both common structural challenges and country-specific opportunities. Based on these findings, we propose a multi-stage regional roadmap for the progressive incorporation of quantum computing concepts into high school education.

It is worth noting that this qualitative research is the first step in a broader research agenda. It sets the contextual foundation for a future quantitative study that will analyze the academic performance, motivation, and learning trajectories of Latin American students participating in quantum education pilot initiatives. Together, these studies aim to pave the way for evidence-based education policy and guide realistic, context-tailored strategies for developing human capital prepared to join and lead the quantum revolution in Latin America.

\section{Background}
The 2025 \textit{International Year of Quantum Science and Technology} represents a milestone that contributes to the consolidation of global initiatives aimed at teaching and learning this discipline \cite{Quantum2025_ES}.  
International organizations have promoted educational programs and strategies to strengthen quantum skills development \cite{OECD2025skills,QTEdu}.  
Technology companies, such as IBM, also contribute with educational platforms that allow students and teachers to access quantum courses and simulators \cite{IBM_QuantumLearning}.

From a traditional educational perspective, the gradual incorporation of content related to quantum computing into high school education does not involve the creation of new subjects, but rather the reinforcement and consolidation of specific topics within already existing educational areas. In particular, the following courses and areas play 
a key educational role:

\begin{itemize}

    \item \textbf{Mathematics}
    \begin{itemize}
        \item \textbf{Complex numbers:} essential to represent the amplitudes and phases of quantum states. Phase information, encoded in the complex part, is essential for describing quantum phenomena such as interference and superposition, and is a central element in the standard mathematical formulation of quantum theory.

        \item \textbf{Linear algebra:} provides the formal language for quantum computing, allowing quantum states to be represented as vectors, logic gates as matrices, and measurement processes as linear operators.

        \item \textbf{Probability and statistics:} provides the basis for the probabilistic interpretation of quantum measurement, wave function collapse, and the non-deterministic nature of experimental results.
    \end{itemize}

    \item \textbf{Physics}
    \begin{itemize}
        \item \textbf{Modern physics:} introduces essential concepts such as quantization, wave-particle duality, and superposition, which make up the conceptual basis for understanding the behavior of quantum systems.
    \end{itemize}

    \item \textbf{Computing}
    \begin{itemize}
        \item \textbf{Computational thinking:} allows abstraction, decomposition, and modeling of systems, as well as introducing the concept of algorithmic complexity. These skills are key to understanding quantum algorithms and their advantages in some cases over classical algorithms.

        \item \textbf{Fundamentals of programming:} allow the development of basic quantum circuits and algorithms using simulation languages and platforms, promoting hands-on and inquiry-based learning.
    \end{itemize}

\end{itemize}

These subjects and conceptual topics are widely recognized academic prerequisites for understanding quantum computing and serve as the basis for the curriculum coverage indicators evaluated later in the methodology and comparative results by country.

\section{Related Work}
Recent literature shows an increased interest in including quantum computing on the high school curriculum, however most studies have been conducted outside Latin America.

\subsection{Specialized courses}
Sun et al. (2024) developed a quantum computing course for high school students in Hong Kong, which integrates quantum simulators, quantum mechanics theory, and programming fundamentals \cite{Sun2024HK}.  The study showed improvements in conceptual understanding, although these were limited in highly specialized contexts.

\subsection{Global curriculum frameworks}
Gragera-Garcés, Gómez-Orzechowski, and Rodríguez-Hernández (2025) proposed a modular framework for incorporating quantum content into the high school curriculum, with an emphasis on interdisciplinarity and progressive learning \cite{gragera2025}.  
This work highlights the importance of adapting resources and methodologies to local realities.

\subsection{Interdisciplinary approaches}
Several authors highlight the importance of linking quantum education to other STEM areas and developing critical and computational thinking as a gateway to quantum concepts.

\subsection{Synthesis}
Taken together, these studies demonstrate a growing enthusiasm for the creation of innovative pedagogical models for teaching quantum computing.
 
Nonetheless, they also reveal a lack of proposals adapted to the Latin American context, where the structural challenges of education systems demand differentiated approaches and contextualized strategies.

Within this context, this research analyzes the challenges, and opportunities for the teaching of quantum computing in Latin America, comparing it with initiatives in developed countries, identifying open source resources and transferable expertise, and proposing a regional roadmap to facilitate the progressive incorporation of this discipline for both teachers and students. 

\section{Latin American Educational Context}

The analysis of national curricula in the selected countries (Peru, Chile, Argentina, Brazil, Colombia, and Bolivia) reveals no significant differences between public and private education in terms of basic content in linear algebra, physics, functions, statistics, or complex numbers, which are commonly included in both sectors.
\cite{UNESCO_STEM2022}.

One important aspect to consider is the disparity in the educational level across countries. In the PISA test results, Chile outperforms the other countries in all evaluated areas, reflecting the quality of its education system \cite{OECD_PISA2022}. Moreover, Chile allocates the highest investment in education within the analyzed group, reaching approximately 5,63\% of its GDP in 2020. \cite{UNESCO_Finance2023}.

However, there are significant differences regarding computer science education. In the countries mentioned, training in this field is usually limited to very basic concepts; in some cases, even elementary programming (such as Scratch) is not covered. In addition, limitations in access to technological resources persist: for example, in Peru, according to data from the Ministry of Education, there is an average of seven students per computer in public high schools \cite{gestion2024}.

A remarkable feature found in the Bolivian curriculum is the extensive coverage of topics such as logic, circuits, electronics, programming, and robotics \cite{MINEDU_Bolivia2010}. However, academic performance remains low: a wide curriculum does not necessarily translate into high academic achievement, possibly due to deficiencies in teacher's preparation, flaws in curriculum design, or, as one teacher interviewed pointed out, national curricula that tends to mask or overstate the current state of the country’s education system (personal interview, 2024). It should be noted that Bolivia does not participate in PISA tests due to the Avelino Siñani–Elizardo Pérez Law (2010), which makes it difficult to compare it to other countries using international standards \cite{MINEDU_Bolivia2010,OECD_PISA2022}. According to the preliminary diagnostic mathematics test for sixth grade (2023), only 3 out of every 100 students achieved a passing grade \cite{UNFPA_Bolivia2023}.

In contrast, countries such as Chile and Argentina have made progress in building computational thinking through approaches focused on competencies and learning hubs \cite{MINEDU_Chile2020,MINEDU_Argentina2022}. There are also prestigious schools —such as ORT, Carlos Pellegrini, Lenguas Vivas, and ILSE— some of which are public and offer International Baccalaureate (IB) programs. \cite{IBO_2024}. In Brazil, the national university entrance exam (ENEM) includes questions on linear algebra, which motivates certain schools to teach matrices and vectors \cite{INEP_ENEM2023}. In addition, advanced programming programs operate in special institutions such as the Núcleos de Altas Habilidades/Superdotação (NAAH/S) (High Ability/Gifted Centers), military schools, and federal institutes \cite{MEC_Brasil2022}. It is worth noting that Brazil invests approximately 6.2\% of its GDP in education, compared to 4.5\% in Peru and Colombia \cite{WorldBank2025,UNESCO_Finance2023}.

In general, most of the curricula reviewed do not include complex numbers, which are fundamental for understanding quantum states. \cite{UNESCO_STEM2022}. Similarly, it would be beneficial to incorporate at least one general-purpose programming language, such as Python, which is currently only taught in so-called “super schools” \cite{GarciaPenalvo2018}.

\section{Metodology}

This study adopts a qualitative and comparative approach, focusing on the analysis of the level of academic readiness in six Latin American countries: Peru, Bolivia, Chile, Argentina, Brazil, and Colombia, to incorporate quantum computing content into high school education. It is based on the assumption that readiness does not depend solely on the formal curriculum, but also on structural, institutional, and sociocultural factors that shape the teaching of STEM disciplines.

This analysis combines a review of educational policies, official curricula, and national programs as well as the identification of regional enablers, as shown below.

\subsection{Conceptual framework}

\textbf{Academic readiness} is defined as the degree of development of the academic, institutional, and contextual competences required to integrate quantum computing into high school curricula.

Academic readiness comprises two main components:
\begin{itemize}
\item \textbf{Academic coursework}:
It refers to the robustness of the curriculum, considering the presence of content on linear algebra, probability, modern physics, computational thinking, and similar skills. It also takes into account the existence of differentiated educational pathways that provide access to this content. (IB, AP  or \textit{elite schools}). 

\item \textbf{Enablers}:
 These are factors external to the school curriculum that influence the feasibility of the implementation, such as infrastructure, teacher training and the availability of open educational resources in the local language.
\end{itemize}

\begin{table}[htbp]
\caption{Description and Assessment indicators}
\label{tab:readiness}
\renewcommand{\arraystretch}{1.3}
\setlength{\tabcolsep}{3pt}

\begin{tabular}{|p{2cm}|p{2.9cm}|p{3.3cm}|}
\hline
\textbf{Category} & \textbf{Description} & \textbf{Assessment indicators} \\
\hline

\multirowcell{2}[-12ex][c]{\textbf{Academic} \\ \textbf{coursework}}
& Curricular structure and academic content in high school education.
& Coverage and depth of key topics: linear algebra, complex numbers, probability/statistics, modern physics, and programming fundamentals. \\
\cline{2-3}
& Alternative learning pathways within the education system. 
& Existence of complementary educational pathways (elective modules, IB/AP programs, high-performance schools, or hubs for scientific talent). \\
\hline

\multirowcell{6}[-27ex][c]{\textbf{Enablers}} 
& Teacher training 
& Existence of professional development programs for teachers in STEM or emerging technologies. \\
\cline{2-3}
& Technology infrastructure 
& Percentage of institutions with internet connectivity and access to digital laboratories or devices; school digitization policies. \\
\cline{2-3}
& Institutions and public policies. 
& Nationwide science, technology, and innovation (STI) strategies, university-school partnerships, and scientific hubs specifically focused on quantum technologies.
\\
\cline{2-3}
& Socioeconomic factors. 
& Equal access to technological infrastructure; urban–rural disparity; public investment per student. \\
\cline{2-3}
& Student culture and engagement. 
& Availability and continuity of STEM clubs, early-stage learning programs, Olympiads, science fairs, or youth communities that promote scientific literacy. \\
\hline
\end{tabular}
\end{table}

\subsection{Scoring system}

To evaluate the level of academic readiness in each country, a five-level ordinal scale (1–5) was established and applied to each assessment indicator described in Table~\ref{tab:readiness}. 

\begin{itemize}
    \item \textbf{1 - Very low:} Absence or minimal presence of the indicator. There are no identifiable policies or programs related to the aspect evaluated.
    \item \textbf{2 - Low:} Early-stage initiatives or regulatory frameworks exist, but with limited or non-verifiable implementation.
    \item \textbf{3 - Moderate:} Formal presence of policies, programs, or curriculum content, albeit with partial scope or uneven coverage.
    \item \textbf{4 - High:} Broad or consolidated implementation of the items assessed, with visible results or institutional monitoring mechanisms.
    \item \textbf{5 - Very high:} Complete and sustainable integration of the indicator into the educational system, supported by active national policies and evidence of impact.
\end{itemize}

\subsection{Assessment procedure}

The assessment process was conducted in three stages:

\begin{enumerate}
    \item \textbf{Documentary analysis:} Official curriculum plans, government documents, and national reports from the six countries analyzed were collected. In cases where official documents were unavailable, priority was given to verifiable sources from international organizations (UNESCO, Economic Commission for Latin America and the Caribbean [ECLAC], Inter-American Development Bank [IDB]).
    
    \item \textbf{Coding and score assignment:} Each criterion in Table~\ref{tab:readiness} was assessed according to the defined scale. Score assignment was agreed upon by the research team based on the analysis of documentary evidence, ensuring consistency across countries.
    
    \item \textbf{Regional comparison and summary:} Comparative matrices were developed by country and dimension, enabling the identification of regional patterns, gaps, and strengths. The results will be presented graphically in the results section using radar charts and heat maps.
\end{enumerate}

This procedure seeks to maintain methodological consistency, transparency, and accountability in decisions made during qualitative coding.

\section{Results}
Following the steps described in the previous section, six countries were studied: Peru, Bolivia, Chile, Argentina, Brazil, and Colombia. The following results were found:

\subsection{Peru}

Table~\ref{tab:peru} summarizes the evaluation of Peru's academic readiness according to the eight indicators in Table~I. The analysis shows progress in basic curricular coverage and high-performance schools (COAR), but also notable limitations in other aspects.

\begin{table}[t]
\centering
\caption{Evaluation of \textit{academic readiness} in Peru (2025)}
\label{tab:peru}
\renewcommand{\arraystretch}{1.35}
\begin{tabular}{p{1.5cm} p{5.3cm} c}
\hline
\textbf{Indicator} & \textbf{Evidence / Description} & \textbf{Score (1--5)} \\ \hline

\textbf{Curricular structure and academic content} &
The National Curriculum incorporates mathematics, physics, and statistics, but lacks formal modules on linear algebra, complex numbers, and modern physics. The programming focus is usually limited to office tools or Scratch. \cite{programa_curricular_2016, minedu_cneb}. &
\textbf{2} \\

\textbf{Alternative educational pathways} &
There are \emph{High-Performance Schools (COAR)} and private schools with international IB and AP programs. 89 schools offering IB have been recorded, making Peru the country with the largest and most decentralized offer among those considered in this study, and 3 schools offering AP. \cite{minedu_coar, ibo_country_peru, collegeboard_apledger_peru}. &
\textbf{4} \\

\textbf{Teacher training} &
Peru has national teacher professional development programs, such as the \textit{Teacher development Program in Pedagogical and Disciplinary Knowledge} and the \textit{Training Program for the Use of Digital Technologies in Pedagogical Practice}, which strengthen pedagogical and digital competencies through the SIFODS platform. These initiatives mainly target general pedagogical improvement and basic ICT use, 
without incorporating advanced STEM or emerging technology training. \cite{minedu_actualizacion_docente_2025, minedu_formacion_tic_2025}. &
\textbf{3} \\

\textbf{Technological infrastructure} &
The 2023 Educational Census shows significant gaps in connectivity and equipment in the Peruvian school system. In public primary schools, 67.3\% of institutions lack internet access, and in public secondary schools, 33.2\% still lack connectivity. Computer access ratios remain high (12 students per device in primary and 7 in high school), with marked urban–rural inequalities. \cite{censoeducativo2023}. &
\textbf{2} \\

\textbf{Institutions and public policies} &
Peru has an active institutional framework in science, technology, and innovation led by CONCYTEC, the Government Secretariat for Digital Transformation (PCM–SGTD), and the Ministry of Education. Recent guidelines include the National Digital Transformation Policy, the draft National Artificial Intelligence Strategy 2026–2030, and the process of developing the National Quantum Technologies Strategy. \cite{concytec_politica_2022, pcm_ia_2026, pcm_cuantica_2025}. &
\textbf{3} \\

\textbf{Socioeconomic factors} &
Public spending on education amounts to 4.5\% of GDP, remaining below the regional average and the lowest among the countries studied. Urban–rural gaps and inequality in access to quality education persist. \cite{worldbank_education_expenditure}. &
\textbf{2} \\

\textbf{Culture and student participation} &
Peru has consolidated initiatives for school scientific participation, such as the national \textit{Eureka} fair and the \textit{National School Mathematics Olympiad (ONEM)} promoted by the Ministry of Education. \cite{eureka_peru, onem_peru}. &
\textbf{3} \\

\hline
\end{tabular}
\end{table}

The average score for Peru is \textbf{2.7/5 (medium level)}. 
The country shows strengths in offering complementary educational pathways and the existence of institutional mechanisms for innovation and digital transformation. 
However, critical gaps in technological infrastructure, limited curricular coverage in advanced content, and the absence of specialized teacher training reduce systemic preparedness to integrate quantum computing into secondary education. 
Nonetheless, the presence of consolidated school scientific communities and the progress of national policies in AI and quantum technologies provide favorable conditions for the development of pilot projects and training strategies led by academic and technological actors.

\subsection{Argentina}

Table~\ref{tab:argentina} summarizes the evaluation of academic readiness in Argentina, which despite its critical economic situation, shows a medium level of readiness and positions itself as one of the countries that invests the most in education in the region. 

\begin{table}[h]
\centering
\caption{Evaluation of \textit{academic readiness} in Argentina (2025)}
\label{tab:argentina}
\renewcommand{\arraystretch}{1.35}
\begin{tabular}{p{1.8cm} p{4.5cm} c}
\hline
\textbf{Indicator} & \textbf{Evidence / Description} & \textbf{Score (1--5)} \\ \hline

\textbf{Curricular structure and academic content} &
The Argentine curriculum provides partial coverage: it includes probability, classical physics, and computational thinking, but lacks advanced content such as linear algebra, complex numbers, or modern physics. There is no national standardization of modern physics. \cite{AR_CurriculoCiencias,AR_CurriculoMatematica} &
\textbf{3} \\

\textbf{Alternative learning pathways} &
Argentina has 56 IB schools, providing access to advanced mathematics and modern physics concepts, although only for a minority of students. \cite{AR_IB_Schools} &
\textbf{4} \\

\textbf{Teacher training} &
The INFoD provides general teacher training; however, it does not guarantee advanced STEM training nor include these areas. \cite{AR_INFOD} &
\textbf{2} \\

\textbf{Technological infrastructure} &
Programs such as Conectar Igualdad aim to provide technological resources; however, in 2022 only 40.7\% of the promised laptops were delivered, and nearly half of public schools lack internet connection. \cite{AR_Infraestructura_TN,AR_Infraestructura_Infobae,AR_ConectarIgualdad} &
\textbf{2} \\

\textbf{Institutions and policies} &
The Ministry of Science and Technology (MINCyT) promotes innovation and scientific outreach, while Argentina's National Council for Scientific and Technical Research (CONICET) develops science fairs and scientific engagement programs, strengthening the STEM ecosystem. \cite{AR_MINCYT,AR_CONICET} &
\textbf{4} \\

\textbf{Socioeconomic factors} &
Argentina exhibits significant socioeconomic disparities and low educational equity, which hinder broad access to advanced STEM education, despite investing approximately 5.2\% of GDP in education. \cite{AR_WB_EducationSpending} &
\textbf{2} \\

\textbf{Student culture and participation} &
Highly developed ecosystem: Science fairs, CONICET, science clubs, and national competitions encourage student participation. \cite{AR_FeriasCiencia,AR_CONICET} &
\textbf{4} \\

\hline
\end{tabular}
\end{table}

The average score of \textbf{3.2/5 (medium level)} indicates that Argentina has a solid foundation in scientific outreach and an adequate availability of educational resources. However, weaknesses in infrastructure and teacher training limit its ability to consolidate as one of the most competitive educational ecosystems in Latin America, restraining the potential that its initial progress could achieve.

\subsection{Colombia}
Table~\ref{tab:colombia} provides an assessment of academic readiness in Colombia.  Colombia shows a growing and comprehensive STEM model, supported by multiple pilot programs executed in small-scale groups. This approach has facilitated territorial expansion into specific rural areas and is expected to operate more efficiently in the future.

\begin{table}[h]
\centering
\caption{Evaluation of \textit{academic readiness} in Colombia (2025)}
\label{tab:colombia}
\renewcommand{\arraystretch}{1.35}
\begin{tabular}{p{1.8cm} p{4.5cm} c}
\hline
\textbf{Indicator} & \textbf{Evidence / Description} & \textbf{Score (1--5)} \\ \hline

\textbf{Curricular structure and academic content} &
The Colombian curriculum is guided by the Ministry of National Education’s (MEN) Basic Competency Standards; it includes competency-based learning outcomes in mathematics, science, technology and complex numbers, but does not incorporate linear algebra or modern physics.\cite{MEN_curriculo_colombia}. &
\textbf{3} \\

\textbf{Alternative learning pathways} &
Colombia has 77 schools authorized by the International Baccalaureate (IB), as well as 23 schools that offer AP.\cite{IBO_colombia_2024}. &
\textbf{3} \\

\textbf{Teacher training} &
Teacher training is based on the MEN's ICT Competencies and the STEM+ approach, which promote interdisciplinary projects and active learning methodologies; urban-rural gaps in access to specialized training persist. \cite{MEN_competenciasTIC_2013, MEN_STEMplus_2022} &
\textbf{3} \\

\textbf{Technology infrastructure} &
More than 21\,000 schools (40\%) lack internet connectivity and nearly 5\,000 do not have electricity, which limits the effective use of ICTs \cite{Infobae_conectividad_2025}. &
\textbf{2} \\

\textbf{Institutions and public policies} &
Colombia has a robust institutional framework (MEN, MinTIC, MinCiencias, National Learning Service [SENA]). National Council for Economic and Social Affairs of Colombia (CONPES) 4069 coordinates policies, financing, and programs to strengthen the educational and technological ecosystem \cite{CONPES_4069}. &
\textbf{3} \\

\textbf{Socioeconomic factors} &
24.1\% of students study in rural areas, where significant gaps in infrastructure, educational quality, and access to STEM resources persist.\cite{DANE_educacion_2023}. &
\textbf{2} \\

\textbf{Student culture and engagement} &
There are science outreach programs, however, state funding is limited and uneven, making it difficult to consolidate an STEM ecosystem \cite{Espectador_recorte_ciencia_2025}. &
\textbf{3} \\

\hline
\end{tabular}

\end{table}

The average score is \textbf{2.7/5 (medium level)}. 
The country shows significant progress in strengthening the STEM approach and expanding scientific outreach initiatives. 
Nevertheless, the lack of sustained public funding and structural limitations in infrastructure —including connectivity and electricity issues affecting thousands of educational institutions— reduce its capacity to consolidate a fully articulated STEM ecosystem, ultimately placing it at a medium level of readiness.

\subsection{Brasil}

Table~\ref{tab:brasil} summarizes the evaluation of academic readiness in Brazil, which shows significant inequalities between regions, hindering the collective development of STEM in the country.

\begin{table}[H]
\centering
\caption{Evaluation of \textit{academic readiness} in Brasil (2025)}
\label{tab:brasil}
\renewcommand{\arraystretch}{1.35}
\begin{tabular}{p{1.8cm} p{4.5cm} c}
\hline
\textbf{Indicator} & \textbf{Evidence/ Description} & \textbf{Score (1–5)} \\ \hline

\textbf{Curricular structure and academic content} &
Brazil possess a comprehensive curriculum established by the Common National Curriculum (BNCC) for basic education; however, this does not apply to modern STEM topics or advanced preparation in quantum computing or modern physics. \cite{bncc_2025} &
\textbf{3} \\

\textbf{Complementary educational pathways} &
Brazil has around 67 IB schools, a moderate number but not comparable to countries like Argentina. However, it has a similar number of schools offering the DP (Diploma Programme). \cite{ib_2025} &
\textbf{3} \\

\textbf{Teacher training} &
Brazil offers specialized teacher training courses provided by the Ministry of Education (MEC), including multimedia resources and lesson plans; nevertheless, these courses are usually not oriented toward advanced STEM. \cite{mec_prof_2025} &
\textbf{2} \\

\textbf{Technological infrastructure} &
According to the TIC Educação 2023 report, between 92\% and 94\% of schools have internet access, and around 90\% have at least one digital device, although only 60\% are available to students, and rural areas remain behind. \cite{tic_2023} &
\textbf{3} \\

\textbf{Institutions and public policies} &
Brazil benefits from an established scientific ecosystem formed by the National Council for Scientific and Technological Development (CNPq), Federal Agency for Support and Evaluation of Graduate Education (CAPES), and Funding Authority for Studies and Projects (FINEP), aimed at promoting research and innovation. The Federal Institutes (IFs) offer STEM courses, although regional inequalities persist. \cite{cnpq_2025, capes_2025, finep_2025, ifs_2025} &
\textbf{4} \\

\textbf{Socioeconomic factors} &
Significant regional inequalities affect educational and technological access, with students in public schools generally having fewer resources. \cite{tic_2023} &
\textbf{3} \\

\textbf{Culture and student participation} &
Brazil has a very active community in science olympiads (OBM, OBF, OBI) and fairs such as FEBRACE, strengthening student scientific culture. \cite{febrace_2025} &
\textbf{4} \\

\hline
\end{tabular}
\end{table}

The average score is \textbf{3.1/5}, reflecting a 
\textbf{medium level} of academic readiness. While the country shows particularly strong sectors —such as its scientific ecosystem and extensive student participation in olympiads— many indicators are weakened by significant regional inequalities. These gaps are especially pronounced in the northern part of the country.

\subsection{Chile}

Table~\ref{tab:chile} summarizes the evaluation of Chile's \textit{academic readiness}. In regional comparative terms, Chile shows a more consolidated curricular foundation and favorable ecosystem, positioning it as a regional benchmark.

\begin{table}[t]
\centering
\caption{Evaluation of \textit{academic readiness} in Chile (2025)}
\label{tab:chile}
\renewcommand{\arraystretch}{1.35}
\begin{tabular}{p{1.5cm} p{5.5cm} c}
\hline
\textbf{Indicator} & \textbf{Evidence / Description} & \textbf{Score (1--5)} \\ \hline

\textbf{Curricular structure and academic content} &
The National Curriculum incorporates complex numbers and probability in mathematics. In physics, it covers modern physics topics. The latter is offered in the differentiated Humanist-Scientific training plan, along with computational thinking and programming. Strong academic performance is reflected in Chile achieving first place in PISA 2022 among Latin American countries. \cite{mineduc_bases_34medio, curriculo_pc_ficha, curriculo_fisica_ficha}. &
\textbf{4} \\

\textbf{Alternative educational pathways} &
The school plan offers scientific-humanistic, technical-professional, and artistic training for 11th and 12th grade students, allowing in-depth study in mathematics up to derivatives, integrals, probability, and statistics; and in technology up to object-oriented programming. Additionally, there are 36 schools with \emph{IB}, no schools offering AP have been recorded. \cite{ibo_country_chile}. &
\textbf{3} \\

\textbf{Teacher training} &
High School science teachers in Chile graduate from university with specialized degrees (\textit{Pedagogy in Physics, Chemistry, Biology, or Natural Sciences}) that combine scientific and pedagogical training, in accordance with Law No. 20.903 establishing the Teacher Professional Development System (2016). 
This structure ensures a solid STEM content foundation. \cite{ley_docente_2016, cpeip_2025}. &
\textbf{4} \\

\textbf{Technological infrastructure} &
Chile has a relatively optimal educational infrastructure thanks to historical policies such as \textit{Enlaces}. Recent initiatives, such as \textit{Connected Classrooms} and programs from the MINEDUC Innovation Center, aim to modernize networks and raise ICT standards in schools. Nevertheless, significant gaps persist between urban and rural areas in connectivity and device availability. \cite{aulas_conectadas,cim_historia}. &
\textbf{3} \\

\textbf{Institutions and public policies} &
Chile has an advanced institutional framework, with an Advisory Commission on Quantum Technologies convened by the Ministry of Science (MINCiencia) in 2024, and the subsequent publication of the national policy document \textit{Estrategia Nacional de Tecnologías Cuánticas 2025--2035} (2025), which defines governance, enabling factors (talent, infrastructure, financing) and adoption guidelines for national sector development \cite{minciencia_estrategia_2025}. &
\textbf{5} \\

\textbf{Socioeconomic factors} &
Chile has one of the highest GDP per capita in the region, which creates favorable economic conditions for access to educational and technological resources. Nevertheless, socioeconomic gaps persist, especially between urban and rural areas. \cite{worldbank_chile}. &
\textbf{3} \\

\textbf{Culture and student participation} &
Chile has a strong scientific ecosystem in schools driven by the MINCiencia Explora Program. Initiatives such as \textit{School Research and Innovation (IIE)} foster inquiry-based projects guided by teachers and specialists, while the \textit{Explora Regional and National Congresses} allow students nationwide to share scientific work. 
\cite{explora_portal}. &
\textbf{4} \\

\hline
\end{tabular}
\end{table}

The average score for Chile is \textbf{3.7/5 (medium–high level)}. The combination of an updated curricular framework, consolidated educational innovation policies, and the existence of a national strategy in quantum technologies makes Chile the most advanced country in the region in terms of institutional organization. The main challenge is territorial equity to ensure homogeneous implementation across all schools nationwide.

\subsection{Bolivia}

Table~\ref{tab:bolivia} summarizes the evaluation of Bolivia's \textit{academic readiness} according to the eight indicators defined in Table~I. The results show limited preparedness, with isolated efforts in infrastructure and STEM culture.

\begin{table}[H]
\centering
\caption{Evaluation of \textit{academic readiness} in Bolivia (2025)}
\label{tab:bolivia}
\renewcommand{\arraystretch}{1.35}
\begin{tabular}{p{1.7cm} p{4.7cm} c}
\hline
\textbf{Indicator} & \textbf{Evidence / Description} & \textbf{Score (1–5)} \\ \hline

\textbf{Curricular structure and academic content} &
The national curriculum includes complex numbers, logic, and electronics, but lacks probability and linear algebra. Moreover, only 3\% of students passed the diagnostic math test and 2\% the diagnostic physics test (2023). \cite{planes_y_programas, unfpa_2023}. &
\textbf{3} \\ 
\textbf{Complementary educational pathways} &
There is the Program for Comprehensive Support to Students with Extraordinary Talent in the Plurinational Education System, but there is no large-scale presence of IB or AP. Only 5 schools offering IB and 6 offering AP have been recorded. \cite{opinion_superdotados_2019, ibo_country_bolivia, collegeboard_apledger_bolivia}. &
\textbf{2} \\ 
\textbf{Teacher training} &
Although Bolivia provides training programs for educators through the Complementary Professional Development Program for In-Service Teachers (PROFOCOM), these focus on strengthening general pedagogical skills under the \textit{sociocommunity productive} model and do not include STEM content for high school teachers. Existing initiatives, such as the \textit{STEM with PhET simulations} course (UMSA–ASDI, 2024), are aimed at university professors. \cite{researchgate_2023, umsa_2024}. &
\textbf{2} \\
\textbf{Technological infrastructure} &
There are initiatives such as “One computer per teacher” and devices loan programs. The 2030 Digital Agenda reports: 85.7\% of urban households have internet access versus only 49.3\% in rural areas, and computer use at home drops from 42.7\% to 10.1\%. \cite{agetic_2023, siteal_computadora_docente}. &
\textbf{3} \\
\textbf{Institutions and public policies} &
Bolivia possess a digital institutional framework led by the Agency for E-Government and Information and Communication Technologies (AGETIC) and the \emph{Digital Agenda 2030}, aimed at expanding connectivity, modernizing public services, and promoting basic digital skills. There are no specific national strategies for cutting-edge sciences or quantum technologies. \cite{agetic_2023}. &
\textbf{2} \\ 
\textbf{Socioeconomic factors} &
High inequality of access and digital divide recognized in national policies. Lowest GDP per capita in South America, but historically the country that invests the most in education (10\% in 2023). \cite{educationprofiles_2023, worldbank_bolivia}. &
\textbf{2} \\
\textbf{Student culture and participation} &
Bolivia has relevant initiatives promoting STEM culture, such as the Plurinational Student Science Olympiads. At the governmental level, the 2030 Digital Agenda includes technological inclusion programs such as \textit{APPventureras}, \textit{RobóTICas}, and the Technology Training and Innovation Centers (CCIT). \cite{minedu_olimpiadas_2024, unifranz_stem_2025, agetic_appventureras_2023, agetic_ccit_2023}. &
\textbf{3} \\
\hline
\end{tabular}
\end{table}

The average score obtained by Bolivia was \textbf{2.4/5 (low–medium level)}. While the country has made progress in expanding ICT initiatives and articulating a national Digital Agenda for 2030, structural limitations persist in teacher training, technological infrastructure, and educational equity. Existing STEM outreach initiatives—such as national science olympiads—indicate a growing interest in scientific literacy. However, the absence of a consolidated national strategy and interinstitutional coordination mechanisms significantly constrains the development of a sustainable and scalable ecosystem for advanced STEM education, including quantum computing.

\section{Centralization and availability of \textit{open source} resources in Latin America}

In this section, we introduce two regional enablers identified through exploratory interviews. These interviews were primarily intended to explore the academic experiences of the participants; however, centralization was repeatedly mentioned as a limiting factor, even though it had not been previously raised by the interviewers. For this reason, we consider centralization to be one of the factors that most strongly influences the enablers discussed in Section~6.

Learning outcomes reveal a persistent urban–rural gap in learning, which reflects the effects of centralization practices. Fernández et al. (2023), based on PISA results, found that urban schools tend to have higher scores than rural schools in all Latin American countries participating in PISA. In addition, they report that the average rural-urban gap in Latin America is 42 points in reading and 33 points in mathematics; this is a very large gap, as it is equivalent to slightly less than half a standard deviation.

In contrast, open-source resources represent a positive cross-cutting enabler due to their availability, which can help mitigate barriers to access. In particular, three types of relevant educational resources can be identified:

\begin{itemize}
    \item \textbf{Open learning platforms for quantum computing:} 
    These platforms offer introductory courses, clear and student-friendly learning materials, and online-accessible simulators, enabling an initial exposure to quantum concepts without requiring specialized infrastructure. A notable example is \textit{IBM Quantum Learning} \cite{IBM_QuantumLearning}.

    \item \textbf{Open-source software libraries and tools:} 
    These resources enable experimentation with quantum circuits and algorithms through classical computing environments, supporting hands-on learning. Widely used software development frameworks (SDKs) include \textit{Qiskit} and \textit{Cirq} \cite{Qiskit, Cirq}.

    \item \textbf{Open educational communities and repositories:} 
    These initiatives facilitate the exchange of learning materials, translations, and pedagogical experiences, contributing to the linguistic and cultural contextualization of content. Emerging regional initiatives such as \textit{QuantumHub Perú} and \textit{QuantumQuipu} exemplify these efforts within the Latin American context \cite{QuantumHubPeru, QuantumQuipu}.
\end{itemize}

While these resources cannot substitute for comprehensive educational policies, teacher training, or technological infrastructure, they do constitute a strategic lever for democratizing initial access to advanced content and supporting pilot interventions in resource-constrained contexts.

\section{Preliminary Roadmap}
\subsection*{Phase 1 (2026--2028): Diagnosis and Initial Training}

Section 6 highlights recurring gaps in specialized teacher training, curricular coverage in advanced mathematics and physics, and unequal access to technological infrastructure, particularly in countries with medium and medium--low levels of readiness. Based on these findings, the first phase of the roadmap focuses on strengthening foundational capacities and enabling minimum conditions for the implementation of pilot initiatives through the following actions:

\begin{itemize}
    \item A Latin American Teacher Development Program in Modern Science and Introductory Quantum Computing, focused on complex numbers, basic linear algebra, probability, and quantum phenomena.
    \item Rapid infrastructure assessments and selection of ``pilot centers'' in each country.
    \item Design of introductory modules that can be integrated into mathematics, physics, and technology courses.
\end{itemize}

\textbf{Country roles:}
\begin{itemize}
    \item Chile leads this phase due to its higher level of readiness, reflected in its overall performance and consistency. Beyond academic capacity, it demonstrates strong institutional coordination.
    \item Peru, building on its strength in complementary educational pathways, leverages COAR schools and high-performance schools as experimental laboratories.
    \item Argentina and Brazil, drawing on the strength of their scientific ecosystems, implement university-supported pilot initiatives.
    \item Colombia initiates urban pilots aligned with existing STEM programs.
    \item Bolivia prioritizes minimum connectivity and foundational teacher training.
\end{itemize}

\subsection*{Phase 2 (2028--2030): Curricular Pilots and Evaluation}

Given that several countries exhibit isolated pockets of educational excellence and complementary pathways available only in a small number of institutions, the findings suggest that the introduction of quantum-related content should begin through controlled and evaluable experiences. This phase proposes targeted curricular pilots to validate pedagogical approaches prior to potential scaling, as follows:

\begin{itemize}
    \item Implementation of curricular pilots in 3--5 schools per country, using open-source resources (e.g., IBM Quantum Learning, Qiskit), in alignment with the regional enablers identified in Section~7.
    \item Frequent academic and motivational evaluation using regionally comparable metrics.
    \item Creation of \textit{school-based quantum clubs} and experimental competitions (Q-Challenges).
\end{itemize}

\textbf{Country roles:}
\begin{itemize}
    \item Chile: Pilots in science-oriented or technical--professional secondary schools, articulated through the Explora IIE program.
    \item Peru: Pilots in COAR schools and additional urban schools.
    \item Argentina/Brazil: Pilots in technical and science-focused schools.
    \item Colombia: Pilots in technical upper-secondary education.
    \item Bolivia: Targeted urban pilots.
\end{itemize}

\subsection*{Phase 3 (2030–2033): Regional Scaling and Ecosystem}

The results also show that, although scientific and institutional capacities exist in some countries, they are fragmented and poorly coordinated at the regional level. Within this context, the third phase aims to consolidate a Latin American ecosystem that coordinates resources, teacher certification, and communities of practice. The following is proposed:

\begin{itemize}
    \item Development of a Latin American repository of open-source resources in Spanish and Portuguese, with translations and validated teaching and learning sequences.
    \item Implementation of a Latin American teacher certification in quantum computing for schools (Q-LATAM), with three levels: mathematical foundations, activity design, and curriculum coordination.
    \item Organization of quantum science fairs and multinational student meetings.
\end{itemize}

\textbf{Country roles:}
\begin{itemize}
    \item Chile, Argentina, and Brazil: Technical and academic reference nodes.
    \item Peru: Educational coordination and standardization through the COAR network and hubs.
    \item Colombia and Bolivia: Expansion of urban coverage and teacher training.
\end{itemize}

\subsection*{Phase 4 (2033--2035): Progressive Curriculum Integration}

Finally, given that most of the countries analyzed do not have systematic curriculum coverage of the fundamental topics for quantum computing, the formal integration of this content is only feasible after the consolidation of prior skills. This phase proposes gradual and contextually adapted curriculum adjustments:

\begin{itemize}
    \item Moderate curriculum adjustments: inclusion of complex numbers, elementary linear algebra, and notions of superposition and qubits in modern physics.
    \item Introduction of abstract reasoning content in national assessments.
    \item Continuity programs for high-achieving students: scholarships, summer courses, and quantum STEM pathways.
\end{itemize}

\subsection*{Summary}

Overall, this roadmap sets out a realistic sequence that starts with the region's current capabilities and culminates in the integration of quantum concepts into the curriculum by 2035. Countries such as Chile, Argentina, and Brazil can lead advanced phases; Peru can play a coordinating role; while Colombia and Bolivia need to strengthen basic conditions before scaling up. Collaboration between ministries, universities, NGOs, and technology hubs will be essential for the sustainability of the process.

\section{Conclusion and Future Work}

This research proposes a comparative and qualitative framework for evaluating the extent to which secondary education systems are prepared to incorporate quantum computing content. The findings show that, although the national curricula of the countries analyzed—Peru, Bolivia, Chile, Argentina, Brazil, and Colombia—include notions of mathematics, physics, and computational thinking that can serve as a basis, the coverage of content fundamental to understanding quantum computing—such as linear algebra, complex numbers, and modern physics—remains limited.
These limitations are reinforced by structural factors, including insufficient specialized teacher training, territorial inequalities in access to technological infrastructure, and the absence—or still early development—of specific national strategies in quantum technologies, with the exception of Chile, which stands out as the country with the highest degree of institutional coordination and government planning in the region.

However, the study also identifies strengths that demonstrate the region's potential to advance toward quantum education. These include the presence of active student communities, high-performance educational programs, consolidated university networks, and growing access to open-source educational resources and digital platforms, which can act as catalysts for pilot initiatives, even in context with limited resources.

Based on this analysis, a multi-stage regional roadmap was designed, emphasizing the gradual and contextualized integration of quantum education. This process begins with teacher training and pilot projects, with the goal of achieving curricular integration by 2035. This roadmap is not intended to be prescriptive, but rather to provide an adaptable guiding framework that enables each nation to move forward according to its level of preparedness, institutional capacities, and educational priorities.

In future research, it is essential to complement this work with quantitative and mixed-method studies that analyze the impact of pilot projects on conceptual learning, motivation, and STEM trajectories, as well as to develop comparable regional metrics to evaluate the results of quantum education in schools. Likewise, the analysis of subnational case studies will allow for a more accurate understanding of territorial dynamics and the proposal of relevant local educational innovation strategies.

This study aims to contribute to the creation of an evidence-based regional agenda for quantum education in Latin America. It focuses on the early adoption of emerging technologies and the strengthening of education systems that are more equitable, integrated, and equipped to address the scientific challenges of the 21st century.

\section{references}

\begingroup
\renewcommand{\section}[2]{}
\footnotesize
\bibliographystyle{IEEEtran}

\endgroup

\end{document}